\documentclass[reprint,notitlepage,onecolumn,prd,11pt]{revtex4-1}

\usepackage{amsmath}

\usepackage{graphicx}
\usepackage{amssymb}
\usepackage{amssymb}  
\usepackage{amsthm}   

\usepackage[x11names]{xcolor}

\usepackage{hyperref}

\hypersetup{colorlinks=false,
    urlbordercolor=LightSkyBlue4,          
    citebordercolor=DarkOliveGreen4,        
    filebordercolor=Firebrick4,      
    linkbordercolor=Firebrick4} %

\begin{document}
\thispagestyle{empty}

\begin{flushleft}
{\Large
\textbf{Collective Phenomena and Non-Finite State Computation in a Human Social System}
}
\\
Simon DeDeo$^{1,2,\ast}$
\\
\bf{1} Santa Fe Institute, Santa Fe, NM 87501, USA
\\
\bf{2} School of Informatics and Computing, Indiana University, Bloomington, IN 47405, USA
\\
$\ast$ E-mail: simon@santafe.edu
\end{flushleft}

\section*{Abstract}
We investigate the computational structure of a paradigmatic example of distributed social interaction: that of the open-source Wikipedia community. We examine the statistical properties of its cooperative behavior, and perform model selection to determine whether this aspect of the system can be described by a finite-state process, or whether reference to an effectively unbounded resource allows for a more parsimonious description. We find strong evidence, in a majority of the most-edited pages, in favor of a collective-state model, where the probability of a ``revert'' action declines as the square root of the number of non-revert actions seen since the last revert. We provide evidence that the emergence of this social counter is driven by collective interaction effects, rather than properties of individual users.

\section*{Introduction}

Social systems---particularly human social systems---process information. From the price-setting functions of free-market economies~\cite{Hayek:1945wp,cottrell:1993cg} to resource management in traditional communities~\cite{lansing2007priests}, and from deliberations in large-scale democracies~\cite{campbell1980american,carpini1997americans} to the formation of opinions and spread of reputational information in organizations~\cite{DeCanio:1998fq} and social groups~\cite{Gluckman:1963vg,Gluckman:1968ul}, it has been recognized that such groups can perform functions analogous to (and often better than) engineered systems. Such functional roles are found in groups in addition to their contingent historical aspects and, when described mathematically, may be compared across cultures and times.

The computational phenomena implicit in social systems are only now, with the advent of large, high-resolution data-sets, coming under systematic, empirical study at large scales. While such studies are well advanced in the case of both human~\cite{chomsky1969aspects,Zimmerer:2010ix} and non-human~\cite{2011PLSCB...7E0011J,tenCate:2012ki} communication, these methods have not been widely applied in the study of collective social behavior.

We study a particular phenomenon, that of cooperation in the online, open source Wikipedia community, with the goal of distinguishing between different classes of computational sophistication. We focus on the distinction between finite and non-finite models, where the latter have access to an effectively unbounded resource, such as a counter, stack or queue~\cite{moore2011nature}.

A feature common to all such analyses is that a finite amount of data \emph{by itself} can never distinguish between two classes whose distinctions are defined in terms of bounded \emph{vs.} unbounded resources. This is sometimes understood in terms of the competence-performance distinction; see Refs.~\cite{chomsky1969aspects} and\cite{de2010grammatical}. Our argument for the emergence of non-finite computational properties thus relies on model selection, and the statistical inference of asymptotic properties of a finite-state system. As part of this argument we prove a result that we refer to as the \emph{probabilistic pumping lemma}: for any finite-state process, and any string $w$, of sufficient length, produced by the process, the probability that a word of length $|w|n$ is found to be $w^n$ decays exponentially as $n$ becomes large. 

The outline of our paper is as follows. We state, and prove, the lemma described above, in the first section, and Appendix~S1. We establish the main empirical result of this work in the second section, where we examine the symbolic dynamics of article editing in Wikipedia. In considering the top ten most-edited articles in the encyclop\ae dia, we find strong evidence in a majority of cases for a violation of the probabilistic pumping lemma, and thus computation over and above that of the finite-state. 

We then discuss the possible origins of this effectively resource-unbounded system in the third section. We conclude with the implications of this finding for the complexity of social systems, and compare our findings with recent work and explore the analogy between formal grammars and social behavior.

\section{The Probabilistic Pumping Lemma}
\label{proof}

In order to distinguish between finite and non-finite models, we focus on the statistics of repeated behavioral patterns, or ``words''. In this section, we show explicitly that probabilistic finite-state process have an exponential cutoff in the asymptotic distribution of repeated words. 

Our discussion here relies on the properties of $P(w^k)$ or, in words, ``the probability of the word $w^k$'', or, more explicitly, ``the probability that a randomly drawn string of length $|w|k$ will be $w^k$.'' Measurement of $P(w^k)$ from data is non-trivial, and detailed discussion of this appears in Appendix~S3.

Our proof establishes the existence of an exponential cutoff by showing that the limiting ratio of $P(w^k)$ (the probability of observing the word $w$ repeated $k$ times in a sample of length $|w|k$), and $P(w^{k+1})$, as $k$ becomes large, approaches a constant strictly between zero and one. We will be able to determine that limiting constant in terms of the properties of the underlying system.

\emph{Statement of Lemma}. For any probabilistic finite-state process, any initial distribution over internal states, and any word $w$, where (1) for all $p$ there exists a $k>p$ such that $P(w^k)>0$ and (2) the system does not deterministically repeat a single word, there exists a positive real number $\epsilon$ such that 
\begin{equation}
\exp{\left[\lim_{k\rightarrow\infty} \sup{\left(\frac{1}{k}\log{P(w^k)}\right)}\right]} = \epsilon,
\label{ppl_formal}
\end{equation}
as $k$ becomes large, with $0<\epsilon<1$, $\epsilon$ \emph{strictly} greater than zero and \emph{strictly} less than one. The limiting value, $\epsilon$, is the spectral radius of $A_{ij}(w)$, the natural extension of the symbol transition matrix to multi-letter words.

The complete proof is given in Appendix~S1. Tests of the numerical convergence of this relation are presented in Appendix~S2, where we study how small machines (number of states of order ten) converge to the bound of Eq.~\ref{ppl_formal} for a uniform prior over spectral radius.

Informally, the lemma says that $P(w^k)$ is bounded above by an exponential cutoff of the form $\epsilon^k$, $0<\epsilon<1$. For most processes, the relevant scale for the limit to obtain is  $k$ of order $p$, the number of states in the underlying process.

Given this, and under the mild assumption that the system has passed through its transient states to one of its aperiodic final classes, the asymptotic probability $P(w^k)$ takes the form of a sum of exponentials,
\begin{equation}
P_\textrm{nEXP}(w^k)=\sum_{i=1}^{n}A_ie^{k\log{\beta_i}},
\label{alpha_simp}
\end{equation}
where here $n$ is the number of classes, and $\beta_i$ are all strictly between zero and one. Eq.~\ref{alpha_simp}, which we refer to as the nEXP model, forms the basis of our model comparisons, and the evidence for non-finite-state computation, presented in the next section.

Note that, for the special case of a purely deterministic (non-probabilistic) machine, where each state has only one transition, either (1) $P(w^k)$ will be zero for all $k$ greater than some fixed value or (2) the output string will just be repetitions of $w$; either violates the conditions of the lemma. Deterministic machines can be recognized by looking for exact repetitions; the more general case that violates Eq.~\ref{alpha_simp}, aperiodicity, can be recognized by non-monotonic behavior.

Note also that the absence of a violation of the probabilistic pumping lemma is not evidence against non-finite-state computation. Even in the case of infinite data, it is easy to construct non-finite-state processes that show exponential decay in all repeated strings; an example can be constructed for a stochastic context-free language that generates strings of matched, but arbitrarily nested, parentheses: ``...()((())())...''.

\section{The Case of Wikipedia}

We now consider a real-world example of collective behavior in a human social system. We are interested in the underlying computational structure of the process, and in particular, the question of whether the system might have access to an unbounded resource. To that end, we compare an infinite-resource model to the general finite-state case using model selection.

\subsection{Model Selection}

A finite-state model, given a sufficient number of states, can reproduce the statistics of an arbitrary process. In statistical study, one must therefore ask when the data justify a simpler (if non-finite) model with fewer parameters. This is known as \emph{model selection}. 

Model selection provides a principled and self-consistent way to select between different descriptions of a process, and to determine (among other things) when adding additional parameters to a model is justified. Without model selection, it would be impossible to establish the existence of a power-law (as opposed to a sum of exponentials), a sine function (as opposed to a finite number of terms in its Taylor series expansion), or a linear trend (as opposed to a truncation of its Fourier decomposition).

Model selection is often done informally, based on the intuitive appeal of one model over another. Here, we attempt a more rigorous approach based on Bayesian methods. The Bayes factor, which provides a self-consistent method for model selection, is now in wide use in the biological~\cite{wade-bayes,suchard-bayes} and physical sciences~\cite{2011PhRvD..83d3505M,2012PhRvD..85j3533E,2012PhRvD..86b3505N,powell-bayes,2012arXiv1209.2024P}. It is of particular use when the question concerns selection between competing hypotheses, rather than (as happens in the frequentist paradigm) the rejection of a null hypothesis~\cite{kass-bayes}. 

For model selection, there are two relevant quantities. The first is $\mathcal{L}$, the log-likelihood of the posterior, or the log of the probability of the data given the best choices of parameters for the model in question,
\begin{equation}
\mathcal{L} = \textrm{log}~{\textrm{max}_{\vec{\alpha}}~{P(D | \vec{w}, M)}},
\end{equation}
where $M$ is a particular model, $\vec{w}$ is the vector of parameters associated with $M$, and $D$ is the data. Models of sufficient generality can, with sufficiently many parameters, make $\mathcal{L}$ arbitrarily large for a given data-set. 

The second quantity, $\mathcal{E}$, is the Bayesian evidence for the model, or, the log-likelihood of the data averaged over all possible parameter values,
\begin{equation}
\mathcal{E} = \textrm{log} \int P(D | \vec{w}, M) P(\vec{w} | M)~d\vec{w}
\end{equation}
It is the Bayesian evidence $\mathcal{E}$ that allows us, in a consistent fashion, to select between models; the reader is referred to Ref.~\cite{mackay2003information}. Meanwhile, the log-likelihood $\mathcal{L}$ is useful as a diagnostic to see which features of the data are relevant. 

The Bayesian evidence requires use of a prior, $P(\vec{w} | M)$; careful specification of the prior is necessary to avoid unfairly penalizing one model over another. In both models we consider, parameters may specify (1) an overall normalization, (2) relative amplitudes of different components, or (3) timescales of decay. We place uniform priors on normalization and decay timescales (within reasonable bounds), and model the priors for relative amplitudes as uniform on the simplex.

To compute $\mathcal{E}$, we use a standard approximation (Ref.~\cite{mackay2003information}; see Appendix~S4). This quantity can be directly interpreted as the log-probability in favor of a model, given the data; thus $\Delta\mathcal{E}$, the difference between $\mathcal{E}$ for two models, corresponds to the log probability in favor of one model versus the other. 

\subsection{Article Timeseries Data}

We consider the ``edit history'' of encyclop\ae dia articles, taken individually. These histories amount to a time-series of editor behaviors: the time-stamped changes to the page made by individuals (either anonymous, or pseudonymous). 

Coarse-graining of these histories is necessary: the number of possible edits that editors can make is essentially unbounded and any edit may change, add, or delete arbitrary amounts of text from the article. A well-known distinction, however, exists between edits that alter the text in a novel fashion and those that ``roll back'' the text to a previous state. The latter kind of edit, called a ``revert'' is used when an editor disagrees with an edit made by someone else and, instead of altering the text further, undoes the work of his or her opponent; as we describe below, revert edits are strongly correlated in time with conflict, and are themselves considered anti-social actions in the context of normal editing.

We thus coarse-grain the history of edits made on an article into two classes, R (``revert'') and C (``cooperate'': any non-revert edit). An example of this process is shown in Table~\ref{sample_history}, while the details of our processing of the raw data are given in Appendix~S3.
\begin{table}
\begin{tabular}{l|l|l|l}
time (UTC) & user & SHA1 (partial) & code \\ \hline 
02:08 & Sarah & {\tt 4abc4aef1ea5 } & \textbf{C} \\ 
05:02 & Alexh25 & {\tt 1e3a2a4656d8 } & \textbf{C} \\ 
05:04 & Mhking & {\tt 4abc4aef1ea5 } & \textbf{R} \\ 
11:39 & Trezatium & {\tt 3b03700b0d9c } & \textbf{C} \\ 
12:15 & Brazilfantoo & {\tt 94a5c05ba10e } & \textbf{C} \\ 
12:31 & Brandon39 & {\tt 3b03700b0d9c } & \textbf{R} \\ 
23:28 & Titoxd & {\tt 109986b8f390 } & \textbf{C} \\ 
23:31 & Titoxd & {\tt 334a315944ce } & \textbf{C} \\ 
23:38 & Titoxd & {\tt 739c15e5bc6a } & \textbf{C} \\ 
23:40 & Titoxd & {\tt 3063a0289680 } & \textbf{C} \\ 
23:42 & Titoxd & {\tt 7aafc8f3f762 } & \textbf{C}
\end{tabular}
\caption{A day of edits on the {\tt George\_W.\_Bush} page, starting at midnight UTC, 21 March 2006. As can be seen by comparing SHA1 hashes of the page content, user Mhking reverted an edit by user Alexh25 to the previous version by user Sarah. Later in the day, user Brandon39 reverted user Brazilfantoo. In between, one can see ``cooperative'' stretches involving both single and multiple users. This sequence of events is coarse-grained into the substring ``CCRCCRCCCCC.'' The full string of (in this case) 45,220 action symbols forms the basis of the finite-state analysis. As with all data used in this study, this sequence is publicly available, in this case at  \href{http://en.wikipedia.org/w/index.php?title=George_W._Bush&offset=200603218&action=history}{{\tt http://en.wikipedia.org/w/index.php?title=George\_W.\_Bush\&offset=200603218\&action=history}} [last accessed 15 August 2013].}
\label{sample_history}
\end{table}

A feature of Wikipedia relevant to this binary classification of edits into revert and non-revert is the presence of so-called ``vandalism''--improper and non-constructive modifications or blanking of the page. Since they usually do not take the form of reversion, these would be classed as C. More detailed descriptions (``prosocial non-revert ''~\emph{vs.} ``antisocial non-revert'') and similarly for the revert case, where pro-social reverts repair vandalism, are certainly possible, and, from the point of view of a detailed understanding, desirable. 

At a coarse-grained level, however, revert edits are a natural class to consider in a study of online conflict~\cite{Suh:2007cb,Kittur:2007ey,Brandes:2008fo}. As noted by Ref.~\cite{Kittur:2010ic}, who studied reversion as a measure of conflict across multiple Wikipedia-like systems, reversions capture implicit cases of task conflict, which are strongly associated with the broader phenomenon of relationship conflict~\cite{DeDreu:2003gz}. Within the Wikipedia community itself, reverts are considered signs of conflict~\cite{reagle}, as can be seen in widely accepted social norms such as the ``three revert rule'' that encourage editors to find ways of resolving conflicts, rather than undoing each other's edits~\cite{cosley2006helping}.

We focus on the most-edited pages, since these provide the greatest amount of data and allow for the most detailed distinctions to be made between pages. While there are large numbers of much less-edited pages, we believe that more sophisticated statistical methods would be required to aggregate this data in such as way as to make statistical study at this level possible.

\subsection{Two Models}

We consider two conceptually distinct models.

The first model is finite; in particular, we consider a finite-state model class of sufficient generality---the probabilistic finite-state machine---that it contains every other model on the finite side of the finite-infinite divide of the computational hierarchy. We consider the probability of seeing an unbroken run of $k$ cooperative events, $C^k$, given that we have just seen a revert, $R$. By the probabilistic pumping lemma, it has the asymptotic form
\begin{equation}
\label{nexp}
P_\textrm{nEXP}(C^k|R) = \sum_{i=1}^n A_i e^{-b_i k},
\end{equation}
where $A_i$ and $b_i$ are free parameters that specify the amplitude and decay rate (timescale) of the $i$th independent component, and $n$ specifies the number of components. 

The second model we refer to as the \emph{collective state} model. In this model, the probability of an additional cooperative event, C, has a functional dependence on the number of cooperative events seen preceding. It is easiest to formulate as the probability of an unbroken run of length $k$,
\begin{equation}
P_\textrm{CS}(C^k|R) = A\prod_{i=1}^k \left(1-\frac{p}{i^\alpha}\right).
\label{asy}
\end{equation}
In words, the collective state model allows for increasing ``returns to scale'': as the number of cooperative events increases, the probability of a non-cooperative event declines as a power-law with index $\alpha$.

Underlying mechanisms have a natural description in the collective state model. In particular, the probability of seeing a \emph{non}-cooperative action, conditional on already having seen $k-1$ cooperative actions just previously,
\begin{equation}
1-P_\textrm{CS}(C|C^{k-1}R) = 1-\frac{P_\textrm{CS}(C^k|R)}{P_\textrm{CS}(C^{k-1}|R)} = \frac{p}{k^\alpha}.
\label{decay}
\end{equation}
scales as a power-law with index $\alpha$. For example, if $\alpha$ is close to unity, then, the collective state model says that the probability of a non-cooperative action declines linearly with the amount of cooperation seen previously. The particular values of $\alpha$ found in the data thus have a direct interpretation in terms of potential underlying mechanisms.

As is clear from Eq.~\ref{asy}, the collective state model violates the probabilistic pumping lemma. It is thus, formally, non-finite. Intuitively, the state space of this model is an effectively unbounded counter that increments with each cooperative event, and resets with each revert.

\subsection{Results}

\begin{figure}
\includegraphics[width=4.5in]{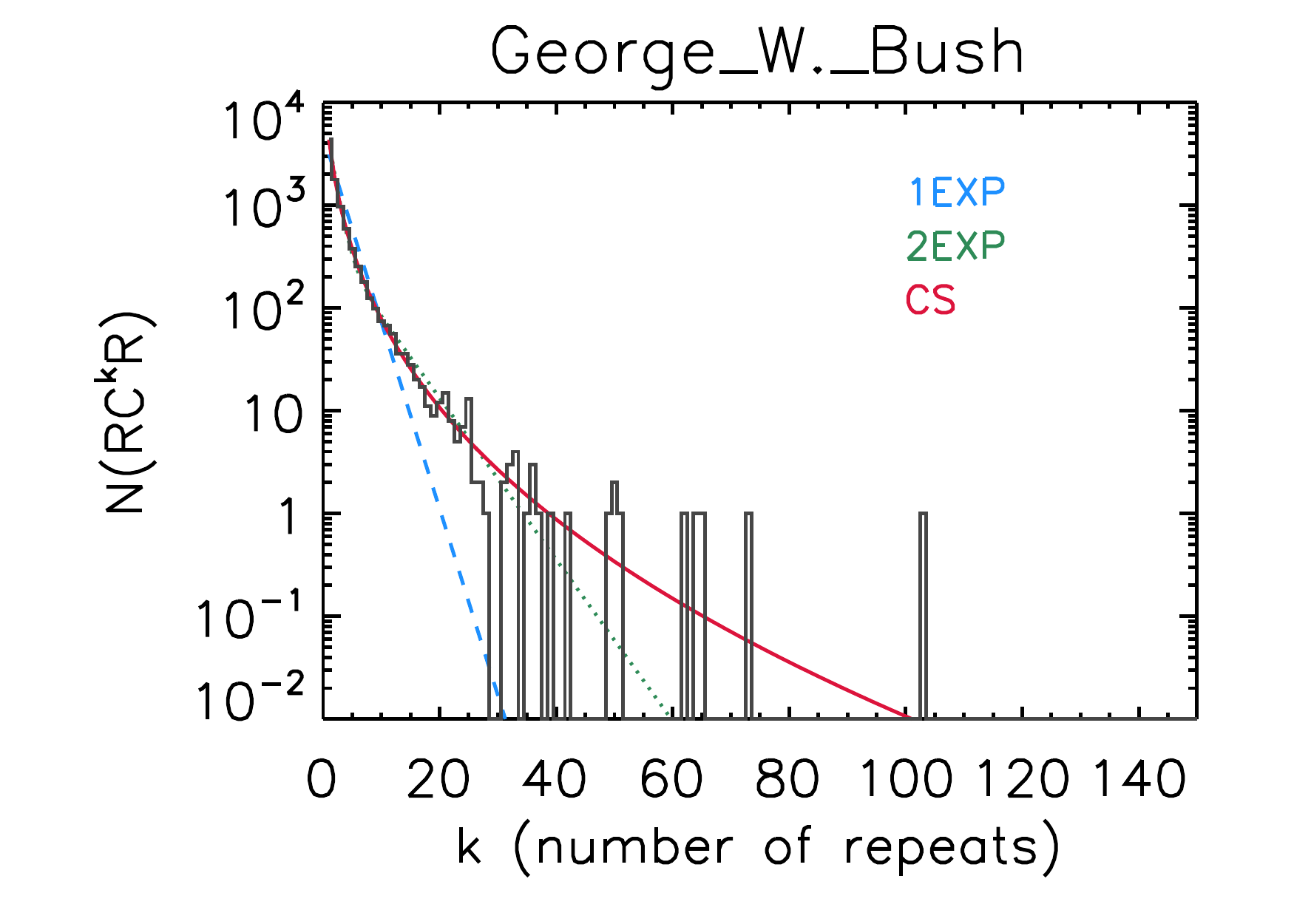} 
\includegraphics[width=4.5in]{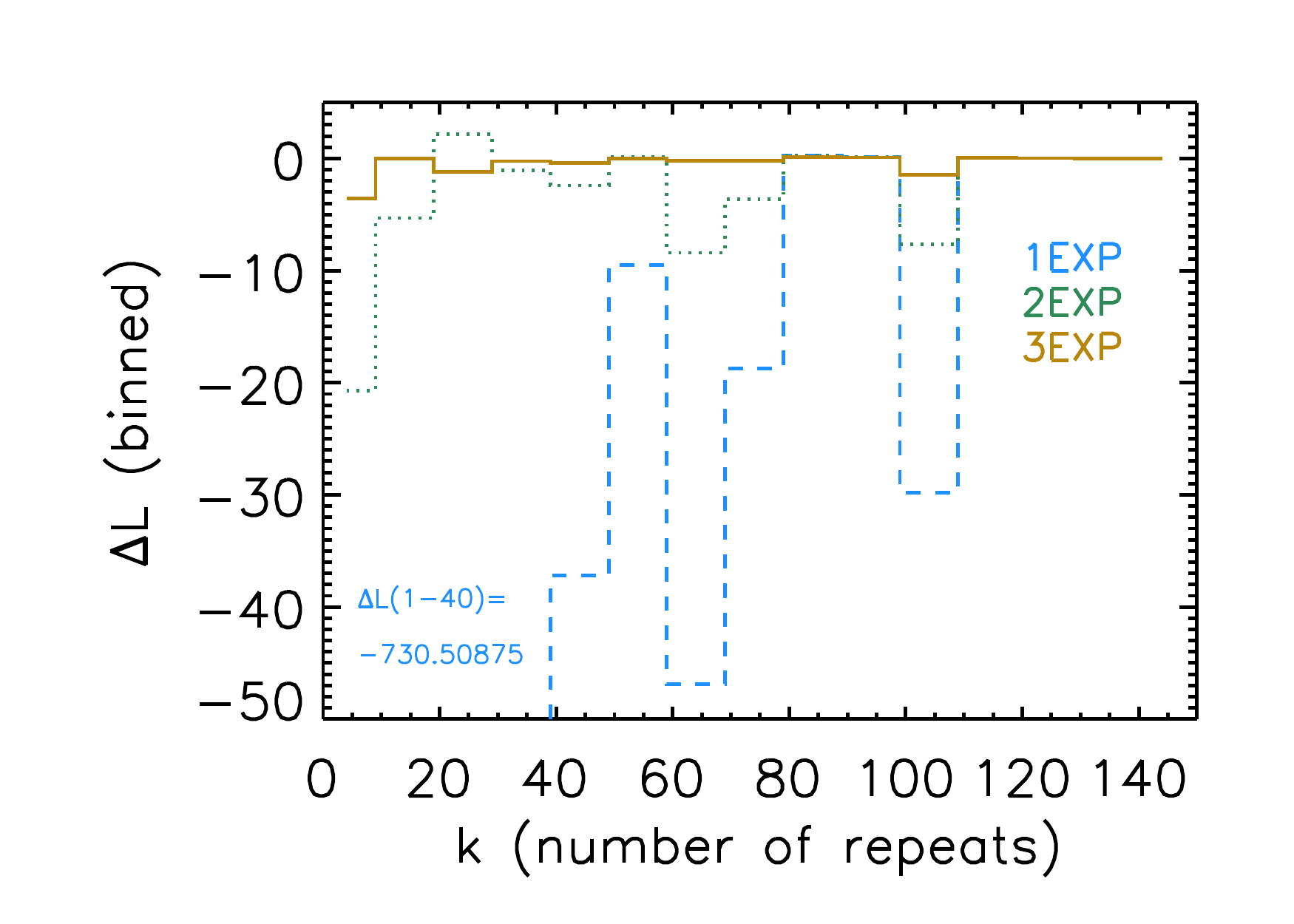} 
\caption{{\bf Top}. Distribution of consecutive C (``cooperative'') events in the edit history of the most-edited article on the English-language Wikipedia, {\tt George\_W.\_Bush}. Solid histogram: actual data. Red/solid line: maximum-likelihood fit for the three-parameter collective state (CS) model of Eq.~\ref{asy}, preferred over the sum of exponential model (nEXP) of Eq.~\ref{alpha_simp}. The blue/dashed and green/dotted lines show the one and two component finite-state approximations to the Collective State model. The finite state model approximates the collective state model in this data at four components (eight parameters), at which point it is strongly disfavored as non-parsimonious by Bayesian model selection. {\bf  Bottom}. Contributions to $\Delta\mathcal{L}$ (log-likelihood relative to collective state) for the one, two, and three component fits (blue/dashed, green/dotted and yellow/solid, respectively).}
\label{gwb}
\end{figure}

Fig.~\ref{gwb} shows the distribution of consecutive C edits for the most edited article in the Wikipedia ``main space'' (\emph{i.e.}, that set of pages supposed to constitute the encyclop\ae dic content): that referring to George W. Bush, the 43rd President of the United States. We refer the reader to Appendix~S3, where we show that counts of the number of strings of the form $RC^kR$, written $N(RC^kR)$, is the preferred data to estimate from.

Even at a glance it is clear that a single exponential---which would appear as a straight line on a log-linear plot---is insufficient to describe the decay of $P(RC^kR)$ as a function of $k$. However, visual inspection alone is insufficient to determine whether to prefer a sum of exponentials (Eq.~\ref{alpha_simp}) to an explicitly non-finite-state process, and we present in Table~\ref{evidence-table} the log evidence ratio, $\Delta\mathcal{E}$, in favor of the collective state model. This table shows that strong evidence against the nEXP model, and in favor of the collective state model, can be found in a majority of cases of the top-ten most-edited articles on the encyclop\ae dia.

\clearpage

\begin{table}
\begin{tabular}{l|l|l|l|l}
sig. & page name & history length & $\Delta\mathcal{E}$ & collective state index \\  
 & & & CS \emph{vs.} nEXP & $\alpha$ \\ \hline 
$<10^{-8}$ & {\tt George\_W.\_Bush} & 45,220 & 18.5 & $0.576\pm0.005$ \\ 
$<10^{-6}$ & {\tt Islam} & 18,054 & 14.9 & $0.592\pm0.007$ \\ 
$<10^{-5}$ & {\tt United\_States} & 31,919 & 12.3 & $0.545\pm0.006$ \\ 
 & {\tt Global\_warming} & 19,541 & 12.1 & $0.602\pm0.008$ \\ 
$<10^{-4}$ & {\tt Wikipedia} & 31,927 & 11.3 & $0.638\pm0.006$ \\ 
 & {\tt Michael\_Jackson} & 26,977 & 10.4 & $0.572\pm0.007$ \\ 
$<10^{-3}$ & {\tt 2006\_Lebanon\_War} & 19,656 & 9.1 & $0.49\pm0.01$ \\ 
 & {\tt Deaths\_in\_2009} & 20,902 & 7.7 & $0.42\pm0.01$ \\ 
$>10^{4}$ & {\tt Deaths\_in\_2007} & 18,215 & -11.5 & --- \\ 
$>10^{7}$ & {\tt Deaths\_in\_2008} & 19,072 & -17.5 & --- \\ \hline 
\end{tabular}
 \caption{log-Evidence ($\Delta\mathcal{E}$) ratios, for the collective state versus the finite-state case, for the ten most-edited pages on Wikipedia. In cases where the collective state model is strongly favored (large, positive $\Delta\mathcal{E}$), we show the best-fit value of the $\alpha$ parameter (see Eq.~\ref{asy}). Eight pages show strong ($p$-value $\leq 10^{-3}$) evidence for the collective state (CS) model of Eq.~\ref{asy} over and above that for the sum of exponentials (nEXP). The strongest evidence in \emph{favor} of finite-state computation is found for two of the three ``death list'' pages, which collate otherwise unrelated information from other parts of the encyclopedia. Appendix~S4 gives details on the use and computation of $\mathcal{E}$ for model selection.}
 \label{evidence-table}
\end{table}

Table~\ref{evidence-table} also presents the collective state index $\alpha$. We find that, in cases where the data favor the collective state model, this index is between $0.42$ and $0.64$; the average value in the top-ten is $0.55$. Eq.~\ref{decay} allows us to interpret this index in terms of the rate at which non-cooperative actions become less likely. 

Our results thus show that the probability of a cooperative run being terminated by a revert action \emph{declines roughly as the square-root of the number of cooperative events} seen in that run. Whatever the underlying nature of the unbounded resources governing the time-series, they must at least be able to maintain a counter, incremented with each C symbol seen, and reset with each R.

\section{Origins of Memory in the Collective State}

In this section, we conduct additional analyses to determine properties of the system that might give clues to the nature of the underlying process.

The results of the previous section provide strong statistical evidence (odds ratios greater than $10^{3}$) for preferring a non-finite model to an explicit enumeration of timescales. The cases in Table~\ref{evidence-table} for which this is \emph{not} the case are themselves of interest. These articles are of a very different nature: ``death lists,'' collections of single sentences listing the dates of deaths of noteworthy individuals. 

That these cases are better described by the sum-of-exponentials model suggests that the article content is relevant to the emergence of non-finite-state computation. This can be either because the user bases that particular content-types attract make it easier for the resultant system to produce non-finite-state behavior. Or, conversely, it could be that the article content itself leads to non-finite-state editing patterns.

\begin{figure}
\includegraphics[width=4in]{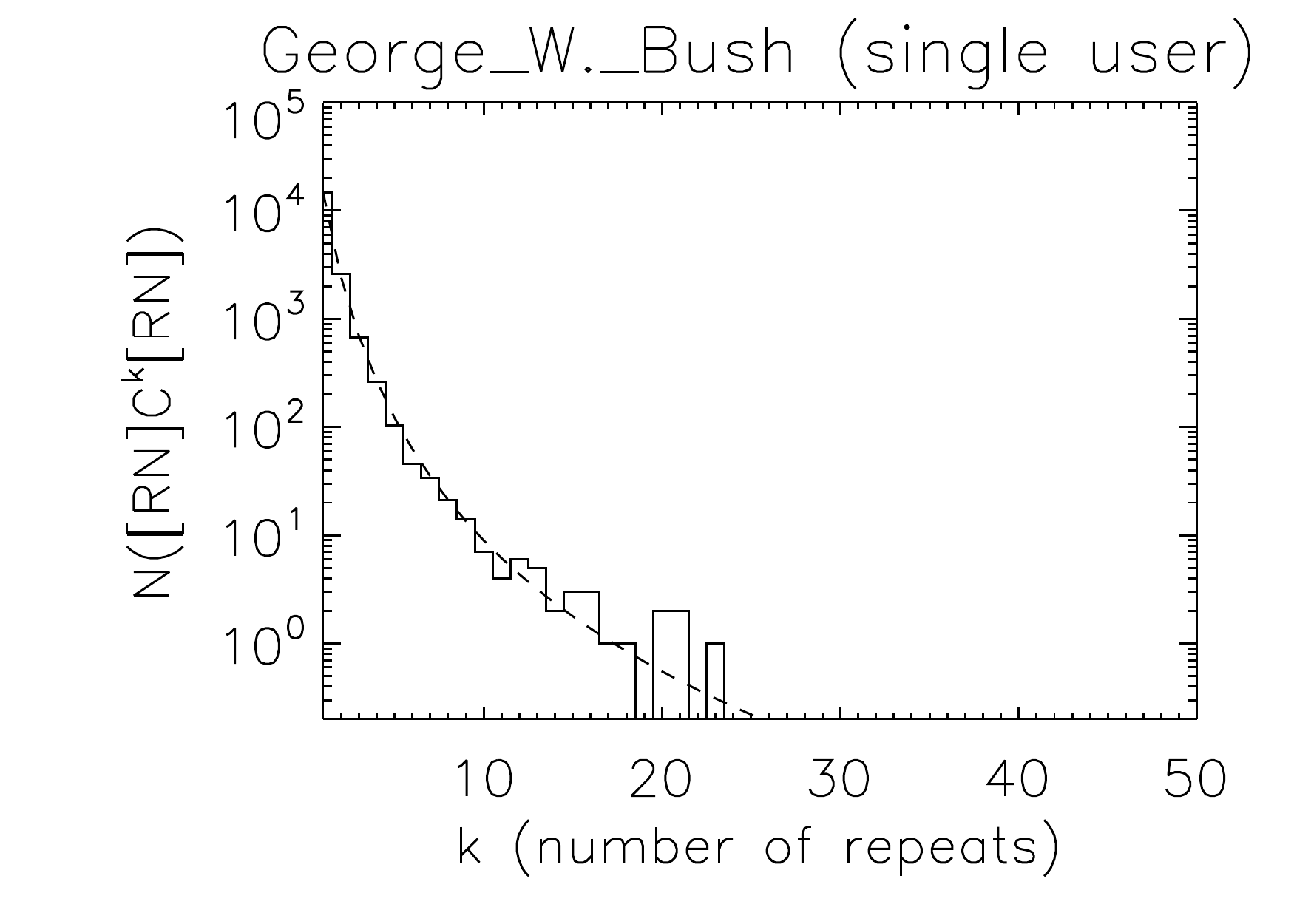}
\caption{Solid line: distribution of consecutive \emph{single-user} C (``cooperative'') events in {\tt George\_W.\_Bush}. The contrast to the multi-user case is clear, showing that long periods of cooperative editing can not be accounted for by unbroken single-user patters. The distribution is well-modeled by the collective state model, Eq.~\ref{asylimit}, with distinct functional form and parameter values from the fit for the multi-user case. The fit is preferred to the finite-state nEXP model at $\Delta\mathcal{E}\approx7.6$ ($p<10^{-3}$).}
\label{gwb-single}
\end{figure}
It could be the case that the cumulative effects associated with the functional form of Eq.~\ref{asy} come from non-interacting users who independently and separately come into contact with an article. The interactions between individuals, on this picture, are unimportant; the content of the page (or a single user's own memory) serves as an effectively unbounded resource that allows violation of the exponential cutoffs required by the finite-state case. 

For example, upon interacting with the page cooperatively, the user might alter it in such a way as to make the probability of a second cooperative edit (by the same user) more likely, and so on. Such a process could potentially lead to behaviors of the same nature as those accounted for by the CS model, without having anything to do with any interpersonal or group-level interaction.

Fig.~\ref{gwb-single} examines this question in detail for the {\tt George\_W.\_Bush} case. We now augment the time-series with an additional symbol, N, representing a change of user (for example, for the data shown in Table~\ref{sample_history}, the new series would be CNCNRNCNCNRNCCCCC), and count strings of consecutive Cs bracketed either by R or N; in other words, a change of user is considered to interrupt the run of Cs. We find the CS model preferred at the $10^{-3}$ level over nEXP; interestingly, the particular functional form of the CS model is the simpler, limiting case
\begin{equation}
P_\textrm{limit-CS}(w^k) = A\prod_{i=2}^k \left(1-\frac{1}{i^\alpha}\right).
\label{asylimit}
\end{equation}
This non-exponential form is not necessarily evidence for non-finite computation in any particular individual; the distribution found for the collection could be understood as the superposition of finite-state machines drawn from a distribution representing the spread of the properties of individuals.

The distinct functional form of the distribution at the individual level suggests that some aspect of interpersonal interaction plays a role in the non-finite nature of the full process. Whether this is driven by how groups are more able to take advantage of the effectively unbounded resource of the page itself (a ``large scratchpad'' model), or because some system memory is encoded in the interactions between the users themselves (an ``interaction combinatorics'' model) is an open question.

An obvious visual difference between Figs.~\ref{gwb} and~\ref{gwb-single} is the elimination of the long tail; it so turns out that long cooperative runs are multi-user events. While it is not the case that long cooperative events necessarily imply the collective state (CS) over the nEXP model (they can be found as well in the ``death list'' pages, where they are fit by a single long timescale exponential component), it is certainly true that the exponential decays implied by the probabilistic pumping lemma require increasingly unlikely fine-tunings of amplitude and decay constants to fit long periods of cooperative behavior.

In the particular case of the George W. Bush page associated with the analysis in this section, the preference for a collective state model in both the individual and the collective case suggests we postulate not one, but at least two distinct counters: one that increments with each C, and is reset with each R, and a second one that increments with each C, and is reset with each R or N.

\section{Conclusions}


This work has examined cooperative behavior in a large-scale social system. We have examined competing models for the processes we observe, and found strong statistical evidence in favor of a \emph{collective state} model. Despite the non-finite nature of the underlying process, the collective state model is more parsimonious than competing finite-state models that approximate it. At the most coarse-grained level of analysis, this model requires at least one ``counter'' that alters the structure of the system over time.

The results comparing collective and individual editing properties further suggest that distinct mechanisms for the violation of the finite-state case are associated with, on the one hand, the cognitive properties of individuals taken separately, and on the other, the fundamentally social phenomenon of Wikipedia as a whole. Distinct counters appear to be running in parallel.

The underlying mechanisms responsible for the emergence of these counters is an open question. They may be fundamentally connected to reputation or memory effects~\cite{Boyd:1989fz,Raub:1990us,Bendor:2001ku}; alternatively, full accounts may require attention to the emergence of social norms~\cite{Ostrom:2000vi, bowles2011cooperative}. Our results here suggest ways to modify and extend ``tit-for-tat'' models of behavior in social systems~\cite{linke2012space} by means of counters that track more fine-grained aspects of system state. 
In addition to these social context effects, the task itself may play a crucial role: the content of the page itself may itself shift the behavior of editors.

This paper has relied on the use of formal languages. First applied to the case of human language~\cite{chomsky1969aspects}, they have now been extended to describe human social interaction (see, \emph{e.g.}, Ref.~\cite{jackendoff2007language} on ``shaking hands''), animal communication~\cite{Hauser:2002p20637,tenCate:2012ki}, animal behavior~\cite{Stoop:2012ux} and pattern recognition more generally (Ref.~\cite{Zimmerer:2010ix} and references therein). This joins the empirical study of cognitive phenomena to a long tradition in the theory of complexity~\cite{Crutchfield199411}. 

When the state of a group is taken to be the sum of the states of the individuals that compose it, coarse-grainings of the system state will in general lead to effective theories~\cite{DeDeo:2011p19975} whose basic units are not descriptions of the state of any one individual. We have previously given such accounts in the case of an animal system~\cite{DeDeo:2010p18133,DeDeo:2011p18663}, where a single formalism is used to attribute computational (``strategic'') states to both individual animals and emergent groups. Ref.~\cite{Flack:2011p20179} provides an explicit analogy between the formal language hierarchy and the decompositions of Ref.~\cite{DeDeo:2010p18133}.

Our work in this paper extends these accounts to human social systems, considered not as ensembles of individual (formal) language users but as a free-standing and unreduced process. Over and above its role in the discussion about cooperative phenomena in social systems, our main result presents a challenge to theory: what formalisms are most natural for the description of non-finite-state processes in the biological and social world? 

Our results demonstrate that empirical study itself can play a role in determining the relative importance of different ways a system can transcend the finite-state aspects of a system: large scratchpads \emph{vs}. interaction combinatorics. While formal language theory presents us with a number of ``post-finite'' languages, such as the context-free grammars and pushdown automata~\cite{moore2011nature}, it seems likely that these will have to be extended or modified to provide tractable models for empirical investigation. 

\section*{Acknowledgements}
I thank John Miller, Nathan Collins, Jim Crutchfield, Ryan James, Cosma Shalizi and Cris Moore, and the attendees of the Santa Fe Institute Complex Systems Summer School 2012, in particular Christa Brelsford, Georg M. Goerg and Oleksandr Ivanov, for helpful conversations. I thank my three referees for careful reading of this manuscript. I acknowledge the support of the Santa Fe Institute Omidyar Postdoctoral Fellowship, the National Science Foundation Grant EF-1137929, ``The Small Number Limit of Biological Information Processing,'' and the Emergent Institutions Project.

\section*{Appendix S1: Proof of the Probabilistic Pumping Lemma}
\label{ppl}

\emph{Statement of Lemma}. For any probabilistic finite-state process, any initial distribution over initial states, and any word $w$, if there exists a $p$ such that for all $k>p$, $P(w^k)>0$~[possibility condition], and the process is not simply a deterministic repetition of a single word $w$, there exists a positive real number $\epsilon$, $0<\epsilon<1$, such that $\exp[\lim_{k\rightarrow\infty} \sup{(1/k) \log{P(w^k)}}]=\epsilon$ as $k$ becomes large. 

\emph{Proof}. We will assume the Mealy machine formalism (observed symbols are emitted upon transitions between internal states~\cite{hartmanis1966algebraic}). Let $A$ be the transition matrix for the process; an element $A_{ij}(\sigma)$ gives the conditional probability of a transition to state $j$, emitting symbol $\sigma$, given that one was previously in state $i$. If the process is reducible, we will assume that sufficient time has passed for the process to reach irreducible subspace of this matrix, and we confine our attention to that subspace. 

We may extend the definition of $A(\sigma)$ to words, as
\begin{equation*}
A_{ij}(w)=\sum_{a_1,\ldots,a_{|w|}} A(w_0)_{ia_1}A(w_1)_{a_1a_2}\cdots A(w_{|w|})_{a_{|w|} j},
\end{equation*}
where $w_i$ is the $i$th symbol in word $w$. We have, further,
\begin{equation}
0<A_{ij}(w) \leq A_{ij}^{|w|}, \label{cond}
\end{equation}
or, in words, the probability to go from state $i$ to state $j$ and emit the word $w$ is less than or equal to that of simply going from $i$ to $j$ in the same number of steps.

By the Perron-Frobenius theorem, the inequality of Eq.~\ref{cond} implies that all eigenvalues, $\beta_i$, of $A_{ij}(w)$ are within the unit circle ($|\beta_i|\leq 1$ for all $i$) with equality obtaining only in the case that $A_{ij}(w)$ is identical to $A^{|w|}_{ij}$. We neglect this latter, trivial case, which only obtains when $w$ is shift-invariant and the all observation runs are given by repeated instances of $w$. Conversely, the possibility condition amounts to the condition that the matrix $A_{ij}(w)$ is not nilpotent, and there exists a non-zero eigenvalue.

If the system (or our knowledge of it) is distributed over its internal states according to probability vector $\pi_i$, we can write the probability of observing a repeated string $w$ as a trace,
\begin{equation}
P(w^k)=\sum^n_{i,j=1} \pi_i A_{ij}^k(w).
\end{equation}
While we have assumed for simplicity that $A_{ij}$ is irreducible, this will not usually be the case for $A_{ij}(w)$. This latter matrix will in general contain both essential and inessential ``self-communicating'' classes\footnote{An index $i$ leads to an index $j$ (written $i\rightarrow j$) iff there exists a $k$ such that $A^k_{ij}(w)>0$. Indices $i$ and $j$ communicate if $i\rightarrow j$ and $j\rightarrow i$. Communication is an equivalence relation, so that classes can be built that contain indices that communicate with each other. Essential classes (sometimes called ``final'' classes~\cite{berman1987nonnegative}) are those which do not lead to any index outside the class; inessential classes are those which may.}
 along with a set of nuisance indices that connect to no other class (\emph{i.e.}, $i$ for which $A_{ij}(w)$ is equal to zero for all $j$)~\cite{seneta2006non}.

The structure of $A_{ij}(w)$ may be visualized as a directed acyclic graph. Inessential classes may have non-zero out-degree, while essential classes, and nuisance indices, are the terminal nodes. Self-loops are permitted, and exist for both inessential and essential classes; these will be crucial to our argument below.

Because the initial distribution $\pi$ may have zero entries, we consider only the part of $A_{ij}(w)$ corresponding to descendants of the non-zero part of $\pi$ in the associated directed acyclic graph. Transitions among the set of nuisance indices, by definition, can not repeat an index. Thus their structure is not relevant to the asymptotic behavior of $P(w^k)$, and we may focus on the essential and inessential classes.

We are particularly interested in the classes that will dominate the $P(w^k)$ probability as $k$ becomes large. Consider the restriction of $A_{ij}(w)$ to a particular class $\alpha$: \emph{i.e.}, construct a submatrix from $A_{ij}(w)$ using only $i,j\in\alpha$. Call this restriction $\alpha_{ij}(w)$. Consider, similarly, the restriction of the distribution $\pi$ to this class. 

Assume first that $\alpha_{ij}(w)$ is diagonalizable. Then, the probability of producing $k$ copies of $w$, while remaining in the class $\alpha$, is
\begin{equation}
P(w^k|\alpha)=\sum_{i,q=1}^{|\alpha|}\beta_q^k\pi^{(q)}v_{i}^{(q)},
\end{equation} 
where $\beta_q$ is the $q$th eigenvalue of $\alpha(w)$, and
\begin{equation}
\pi_i=\sum_{q=1}^{|\alpha|} \pi^{(q)}v_i^{(q)}.
\end{equation}
By construction of the equivalence classes, $\alpha$ is irreducible. Then, by the Perron-Frobenius theorem, the largest eigenvalue of this matrix, $\beta_1$, is real, has a strictly positive eigenvector, and $\pi^{(1)}$ is necessarily greater than zero. 

If $\alpha_{ij}(w)$ is acyclic then $P(w^{k}|\alpha)$ can be written
\begin{equation}
P(w^k|\alpha)=A_1\beta^k_1\left(1+\sum_{i=2}^{\alpha}A_i\left(\frac{\beta_i}{\beta_1}\right)^k\right),
\label{alpha_simp_proof}
\end{equation}
where $A_1>0$, $\beta_1$ is real, and $|\beta_i|<\beta_1$ for all $i>1$, and 
\begin{equation}
\exp{\left[\lim_{k\rightarrow\infty} \left(\frac{1}{k}\log P(w^k|\alpha)\right)\right]}= \beta_1.
\label{alpha_comp_simp}
\end{equation}

If $\alpha_{ij}(w)$ is diagonalizable, but the period, $d$, is greater than one, we will have additional eigenvectors associated with complex rotations of $\beta_1$, $\beta_1\exp{2\pi i k/d}$, $k=\{1\ldots d-1\}$. These will lead to additional oscillatory terms in the leading order term; these oscillations of will be governed by an overall exponentially-decaying envelope, so that 
\begin{equation}
\exp{\left[\lim_{k\rightarrow\infty} \sup{\left(\frac{1}{k}\log P(w^k|\alpha)\right)}\right]}= \beta_1,
\label{alpha_comp}
\end{equation}
regardless of the period of $\alpha_{ij}(w)$.

Finally, consider the case of non-diagonalizable $\alpha_{ij}(w)$. In this case, the matrix can be brought into Jordan normal form, with $m$ blocks, each of size $n_i$ and associated with an eigenvalue $\beta_i$. Assume that the matrix is aperiodic. By the Perron-Frobenius theorem, $n_1$ is equal to one~\cite{meyer2000matrix}. The $k$th power of $\alpha_{ij}(w)$ can then be written (see, \emph{e.g.}, Ref.~\cite{andrieux}),
\begin{equation}
P(w^k|\alpha)=A_1\beta_1^k+\sum_{i=2}^{m} \left(\sum_{j=0}^{n_i-1} A_{ij}{k \choose j}\beta_i^{k-j}\right),
\label{alpha_jordan}
\end{equation}
where $A_1>0$, $\beta_1$ is real, and $|\beta_i|<\beta_1$ for all $i>1$ as before. When $k$ is greater than the largest block size, we can write
\begin{equation}
P(w^k|\alpha)=A_1\beta_1^k\left(1+\sum_{i=2}^{m} f_i(k)\left(\frac{\beta_i}{\beta_1}\right)^{k}\right),
\label{alpha_jordan_asy}
\end{equation}
where $f_i(k)$ is a polynomial function of $k$, of degree $n_i-1$. Eq.~\ref{alpha_jordan} thus obeys Eq.~\ref{alpha_comp_simp}; for a non-aperiodic $\alpha$, an argument identical to the above gives the convergence of Eq.~\ref{alpha_comp}.

Having understood the single-class case, we now consider $w^k$ strings generated by multiple classes. 

Any particular string $w^k$ may be generated by a set of transitions within and between classes. Because these transitions are governed by the directed acyclic graph structure, there will be a finite number of transitions between states. Thus, as $k$ becomes large, the probability of $P(w^k)$ for a particular set of transitions will be governed by the self-transitions, given by terms of the form Eq.~\ref{alpha_comp}. 

In particular, $P(w^k)$ is the sum of a finite number of terms; each term in the sum is a product of at most $p$ transitions between classes, and at least $k-p$ terms of the form $P(w^n|\alpha)$, for different $\alpha$. Explicitly,
\begin{equation}
P(w^k)=\sum_{i\in p(G)} T_i \prod_{j=1}^{N} P(w^{n_{i,j}}|\alpha_j),
\label{sum_of_terms}
\end{equation}
where $i$ indexes the paths of length $k$ through the graph $G$ representing the underlying $A_{ij}(w)$ structure, $T_i$ is a prefactor governing the probabilities of transitions between classes, N is the number of classes, and the total number of within-class transitions is forced to grow with $k$,
\begin{equation}
\sum_{j=1}^N n_{i,j} \geq k-p
\label{sum_part}
\end{equation}
for all possible paths $i$. 

For large $k$, the growth in the number of possible paths (\emph{i.e.}, the growth of the $|p(G)|$) is bounded by the growth in the number of ways to partition the sum in Eq.~\ref{sum_part}. In particular, for large $k$, the number of possible paths relevant to $P(w^k)$ can increase only polynomially in $k$.\footnote{For any set of transitions between classes, the number of self-transitions is bounded by $k$, and the number of distinct classes to assign those self-transitions to is bounded by $p$, the number of states in the machine. The number of ways $p$ terms can sum to $k$ is ${k-1\choose p-1}$, which is bounded by the polynomial $k^p$.}

Meanwhile, each term in the sum of Eq.~\ref{sum_of_terms} is decreasing exponentially, governed by products of the $\beta_{i,1}$, the largest eigenvalues for the classes that have self-transitions for that term. The dominant terms in the sum will be those for which the exponential decline is slowest. By the Perron-Frobenius theorem, the largest eigenvalue of a submatrix associated with a class of $A_{ij}(w)$ is equal to the spectral radius of the matrix as a whole. If $P(w^k)$ is greater than zero for $k$ larger than $p$, the pigeonhole principle invoked in the ordinary pumping lemma~\cite{pippenger1997theories} allows us to assume the existence of at least one self-communicating class; this then means that the spectral radius is equal to that of $A_{ij}(w)$ itself.
\begin{equation}
0<\exp{\left[\lim_{k\rightarrow\infty} \sup{\left(\frac{1}{k}\log{P(w^k)}\right)}\right]} = \rho(A_{ij}(w)) < 1,
\label{proven}
\end{equation}
which was to be proved.\qed

While our paper presents the first explicit application of this form of reasoning to human social systems, we note in passing the use of this kind of reasoning in the study of bird song. Once regarded as strictly finite-state~\cite{Berwick:2011jh}, the sound sequences produced by songbirds are now recognized to show features of non-finite-state computation. A recent, compact model of song production in the Bengalese finch (\emph{Lonchura striata domestica})~\cite{2011PLSCB...7E0011J}, demonstrates the need for a self-modifying (and thus non-finite-state) Markov process. 

An analysis of data on a different species, the Zebra finch (\emph{Taeniopygia guttata}), shows that the probability of an additional repetition, the analog of this paper's $P(C^{k+1})/P(C^k)$, decreases exponentially~\cite{languagelog}. This is, of course, the other way to violate the probabilistic pumping lemma (under the assumption of having reached an aperiodic final class)---the exponential of the lim-sup, Eq.~\ref{alpha_comp}, goes to zero as opposed to unity. It is just as much evidence against finite-state computation, but found in the anomalous absence, rather than presence, of extreme events.

\section*{Appendix S2: Numerical Tests of Convergence Properties}
\label{simulation}

With a view towards determining how the lemma of the previous section applies to actual finite-state processes, we study a restricted class of machines numerically. We sample from the space of probabilistic unifilar machines with $p$ states over a two-symbol alphabet. Such a system can be represented by a weighted, directed graph, with each node having at least one, and at most two outgoing edges, each of which is associated with one of the two symbols, and whose weights sum to unity.

\begin{figure}
\includegraphics[width=4in]{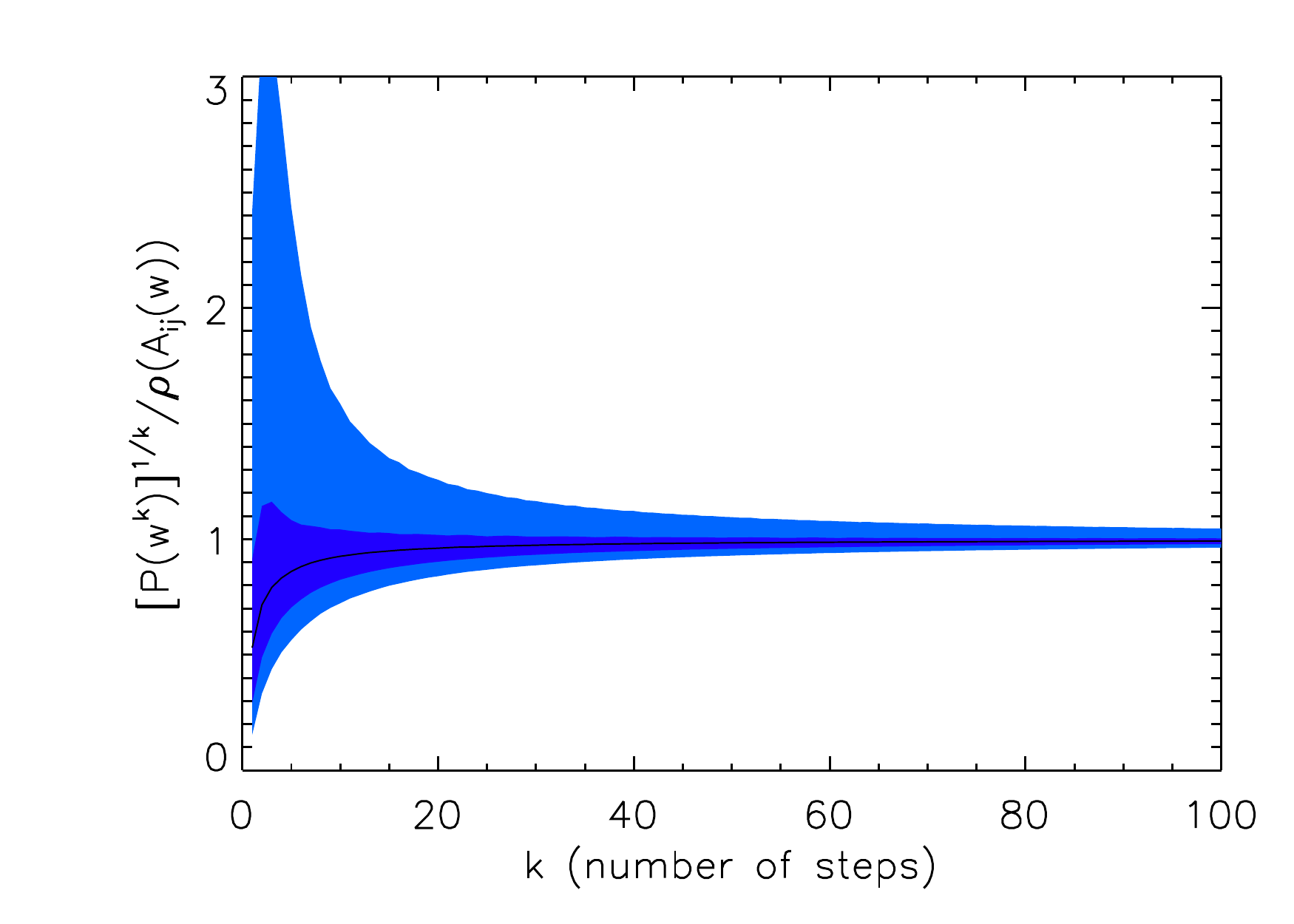}
\caption{Numerical study of convergence of repeated word frequencies to exponential decay, with cutoff predicted by the spectral radius. Shown here is the measured decay rate to the asymptotic limit predicted by Eq.~\ref{proven}, for irreducible finite-state processes with ten states, two output symbols $\{C,R\}$, $w$ equal to C, and a uniform distribution over values of $\rho(A_{ij}(w))$, the spectral radius and asymptotic decay rate, between $0< \rho < 1$. Light blue shows $2\sigma$, and dark blue $1\sigma$ ranges about the median value. For empirical work, convergence is much faster when considering $[P(w^{q+k})/P(w^{q})]^{1/k}$, with $q$ larger than the (assumed) number of states.}
\label{stats}
\end{figure}
For small $p$, the underlying graph-theoretic space can be described completely: for each node, we have a choice of one vs. two outgoing edges; in the case of only one outgoing edge, we must choose between the two symbols. Neglecting the possibility of equivalent machines, we then have the number of such machines, as a function of $p$, as
\begin{equation}
N(p)=(2p+p^2)^p,
\end{equation}
which grows rapidly: there are 12 billion such machines with six states, and more than $10^{400}$ with one hundred states.

We are most interested in how quickly the statistics of an actual machine approaches the limiting value given by Eq.~\ref{proven}. For any particular $A_{ij}(w)$, we can compute the spectral radius and compare that to the ratio $P(w^k)/P(w^{k-1})$ found for distributions over initial conditions that include a self-communicating class as a function of $k$.

\begin{figure}
\includegraphics[width=4in]{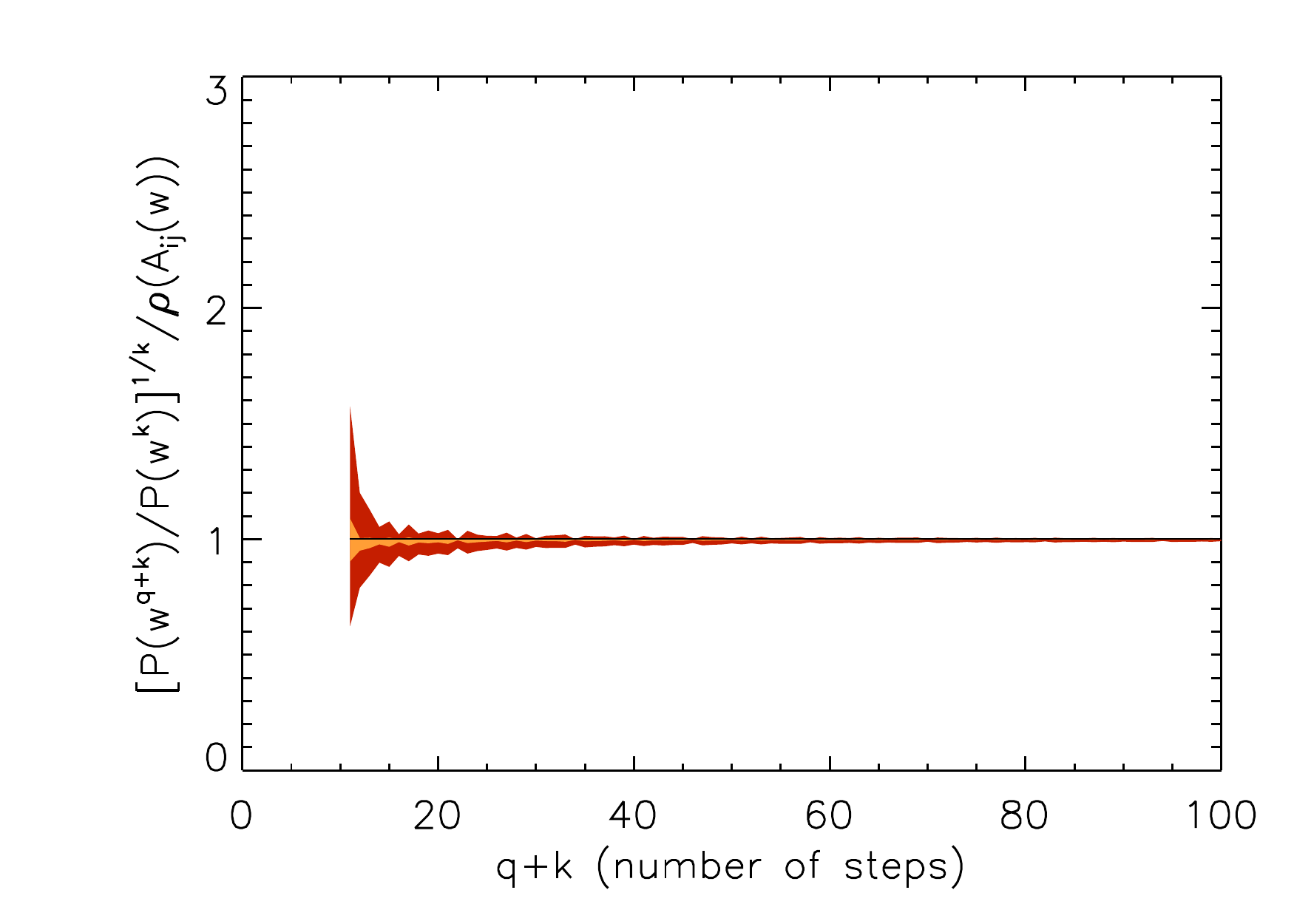}
\caption{Convergence to exponential cutoff as seen with $\hat{C}(q,k)$ (Eq.~\ref{chat}), for the same system as in Fig.~\ref{stats}. Here we take $q$ equal to ten, the number of states. For the same amount of data, convergence is faster for $\hat{C}$ than C; here convergence for $\hat{C}$ to the asymptotic value (at $1\sigma$ confidence), is achieved for $k$ equal to thirty.}
\label{betterstats}
\end{figure}

In Fig.~\ref{stats} we show convergence to the limit by sampling the space of strongly-connected ten-state machines, and considering the frequency of a single repeated symbol. We take a uniform prior over $\rho(A_{i}(w))$, the spectral radius and limit established by the lemma of the previous section, and show the convergence ratio, \emph{i.e.},
\begin{equation}
C(k)=\frac{[P(w^k)]^{1/k}}{\rho(A_{ij}(w))},
\end{equation}
to provide a numerical example of the limiting process established in the previous section. For small $k$, $P(w^k)$ may be dominated by movement through nuisance states and inessential classes, and by contributions from essential classes that have small self-communication probability. Convergence to the spectral radius thus occurs much faster when considering
\begin{equation}
\hat{C}(q,k)=\frac{[P(w^{q+k})/P(w^q)]^{1/k}}{\rho(A_{ij}(w))},
\label{chat}
\end{equation}
where $q$ is longer than the relevant scales of the transient phenomena (\emph{e.g.}, at least as large as the assumed number of states.) This is shown explicitly in Fig.~\ref{betterstats}, where we take $q$ to be the number of states in the system.

\section*{Appendix S3: Details on Coarse-Graining and Analysis of Wikipedia Behavior}
\label{wiki_app}

Our coarse-graining of behaviors on any particular page aims at locating where one user reverts (undoes) the contributions of another editor completely. We locate reversion edits in two distinct ways. Firstly, following Ref.~\cite{Kittur:2010ic}, we can identify reversion edits by the presence of keywords, such as {\tt rv} and {\tt revert},  in the edit summaries; we do so with the following regular expression: {\tt /([Rr][v]+[\textbackslash~\textbackslash n]|[Uu]ndid|[Rr]evert)/}. Secondly, following analyses such as those of Ref.~\cite{Yasseri:2012ga}, we can look for versions of a page with identical SHA1 checksums; the version with the later timestamp may thus be considered a revert to the earlier page. In general, these two metrics align very well, although not perfectly; in this work, we focus on the latter method as a more objective one that does not rely on editors self-reporting. We do not include self-reverts, or edits that do not alter any aspect of the page (\emph{i.e.}, that would otherwise look like ``reverts to the current version'').

The probabilistic pumping lemma works in terms of $P(w^k)$, and our analysis considers the probability of repeated cooperation. However, the measurement of $P(C^k)$ in the data, if done naively, leads to unacceptable results. In particular, estimating $P(C^k)$ for a particular page by counting the number of times the string $C^k$ appears in the time-series, leads to strong bin-to-bin correlations, since an observation of a string $C^k$ necessarily leads to observations of  strings of the form $C^{k-1},C^{k-2},\ldots,C^{k-\lfloor k/2 \rfloor+1}$, and then \emph{two} observations of the form $C^{k-\lfloor k/2 \rfloor}$, and so on. This would lead to excessive complications in the likelihood analysis; conversely, if the correlations are neglected, it leads to claims of heavy-tailed distributions that spuriously rule out exponential decay.

Instead, we count prefix- and suffix-free strings that do not have this shift problem---in particular, we consider the quantity $N(RC^kR)$. As long as $N(RC^kR)$ is significantly less than N, counts of $RC^kR$ and $RC^mR$ are independent of each other and we can write
\begin{equation*}
P(RC^kR)\approx\frac{N(RC^kR)}{N}.
\end{equation*}

The quantity $P(RC^kR)$ itself can be written as
\begin{eqnarray}
P(RC^kR) = P(R)P(C^k|R)P(R|RC^k) & = & P(R)P(C^k|R)\left[1-P(C|RC^k)\right] \nonumber \\
& = & P(R)\left[P(C^k|R)-P(C^{k+1}|R)\right]. \label{deltap}
\end{eqnarray}
In the case that $P(C^k|R)$ is the sum of exponentials in $k$, we have
\begin{equation}
N(RC^kR) \propto P(C^k|R) \propto P(C^k),
\label{derivation}
\end{equation}
or, in words, that if $P(C^k)$ is a sum of exponentials, so is $N(RC^kR)$. The relationship between these two quantities is not always so simple; in the collective state (CS) case, Eq.~\ref{deltap} implies that the quantity $N(RC^kR)$ has a different functional form from $P(C^k)$. In particular, we have
\begin{equation}
P_\textrm{CS}(RC^kR) = \frac{Ap}{(k+1)^\alpha} \prod_{i=1}^k \left(1-\frac{p}{i^\alpha}\right),
\label{dasy}
\end{equation}
which is the functional form we fit and display in Fig.~1 of the Main Article.

\section*{Appendix S4: Details on Model Selection}
\label{stat_app}

In this section we describe in greater detail our methods for distinguishing between the asymptotic and exponential models.

Computation of the likelihood ratio requires an error model for the distributions of $N(RC^kR)$. Since we lack an explicit model for the errors themselves, as a first approximation, we take measurements of $N(RC^kR)$ to be identically and independently distributed. For $N(RC^kR)\ll N$, N the total number of observations, this is a reasonable assumption. Given this, the Poisson distribution of counts follows, and computation of $\mathcal{L}$, the log-Likelihood, or $\log{P(D|\vec{w}, M)}$, for any particular model $M$ with parameters $\vec{w}$, can be written as
\begin{equation}
\Delta\mathcal{L} =  \sum_{k=1}^{k_{\mathrm{max}}} N(RC^kR)\log{\lambda(\vec{w}, k)}-\lambda(\vec{w}, k),
\end{equation}
where we drop model-independent constants. Given sufficiently flat priors, $P(\vec{w}|M)$, around the peak of this function, this is sufficient to estimate many quantities of interest, including the maximum \emph{a posteriori} values of $\vec{w}$ and the error bars on those estimates. 

Our main goal, however, is not parameter estimation, but rather \emph{model selection}, where one compares models with different parameter spaces. In our particular case, one class of models (nEXP) can approximate, by superposition of exponentials, the other class (CS). As the number of exponentials in the sum increases, the approximation becomes increasingly good. We would like to know when we are justified in preferring the more parsimonious model.

Two main frameworks for the resolution of this question exist. On the one hand, the Aikiake Information Criterion (AIC) can be used to estimate the expected KL divergence between the predictions of a model and the true process. In the limit of large amounts of data, it prescribes a constant penalty of $k$, the number of parameters, to the likelihood.

This penalty is sometimes taken as an ``Occam penalty,'' but the correct interpretation is as a guide for prediction out of sample. Prediction out of sample is a conceptually distinct problem, since a complicated approximation to the true model may work very well in a limited range, particularly in the presence of experimental noise. In Monte Carlo testing, AIC tends to prefer complicated approximations, even in cases where the underlying model is more parsimonious~\cite{Kass:1995eh}; a related formal result is that AIC is ``dimensionally inconsistent,'' meaning that even in the limit of infinite data, use of the AIC will lead to non-zero probability of choosing an (incorrect) approximation~\cite{Liddle:2004cz}. 

On the other hand, one can compute (or approximate) what is called the Evidence\footnote{A common rough approximation to this function gives the BIC, or ``Bayesian Information Criterion,'' which prescribes a penalty of $n\log{|D|}$, where $|D|$ is the amount of data.}, which requires knowledge of both the likelihood, $P(D|\vec{w}, M)$, and the prior expectation of parameter ranges, $P(\vec{w}|M)$,
\begin{equation}
E=P(D|M) = \int P(D|\vec{w}, M) P(\vec{w}|M)~d^kw,
\label{evidence}
\end{equation}
where $k$ is the number of parameters (dimensionality of $\vec{w}$). Formally, the Evidence is proportional to ``the probability of the model $M$, given the data observed,'' if equal prior probability is given to the models under consideration. As in all model selection cases, absolute values of the Evidence are irrelevant. One considers only ratios and phrases the question, as in Table~\ref{evidence-long}, as to whether (for example) ``model A is at least a factor of $10^3$ more likely than model B.'' 

In this work, we take the latter approach, operating entirely within the Bayesian framework. This is because our contrasting model classes have small numbers (less than ten) of parameters, all of which have clearly specifiable priors, $P(\vec{w}|M)$. Computation of the full posterior is now common when these circumstances obtain, as is often the case in the exact sciences~\cite{2011PhRvD..83d3505M,2012PhRvD..85j3533E,2012PhRvD..86b3505N}.

In order to calculate $E$, we use the Laplace (or saddle point) approximation; in log-units,
\begin{eqnarray*}
\mathcal{E}=\log{E} \approx \mathcal{L}({\vec{w}_\textrm{max}}) + \log{P(\vec{w}_\textrm{max}|M)} \\
- \frac{1}{2}\log{\det{A}} + \frac{1}{2}k\log{2\pi},
\label{evidence-laplace}
\end{eqnarray*}
where $\mathcal{L}$ is the log-likelihood, $\vec{w}_\textrm{max}$ are the parameters that maximize the likelihood, and $A$ is the Hessian, equal to
\begin{equation}
A_{ij}=\left. -\frac{\partial^2 \mathcal{L}}{\partial w_i \partial w_j}\right|_{\vec{w}_\textrm{max}}.
\end{equation}
We refer the reader to Ref.~\cite{mackay2003information} for details on this approximation.

It remains to specify the priors $P(\vec{w}_\textrm{max}|M)$ for the two models. The nEXP class has $2n$ parameters; the CS class has $3$. The parameters are of two kinds. 

Both nEXP and CS have parameters corresponding to the one-step decay of the underlying quantity $P(C^k)$. In the case of nEXP, there are $n$ such parameters, $b_i$, that play this role. In the case of CS, there is only one, $p$. We take a uniform prior in $p$ (CS) and $b_i$ (nEXP). We allow all $p$ to range independently between zero and $0.995$; the high end corresponds to an exponential cutoff of order $200$ repeats, much longer than seen in the data.

We then have normalizations of terms ($n$ normalizations for nEXP, one for CS). These are fixed by the value of $P(C^1)$, the overall cooperative fraction.
\begin{equation}
N(C)\approx NP(C).
\end{equation}
The maximum value of $P(C)$ is unity. This then leads to an overall area factor of
\begin{equation}
\frac{N^n}{n!},
\end{equation}
for nEXP, where the factor of $n!$ is because the overall sum of all normalizations is confined to the interior of an N-dimensional simplex. In the case of CS, $P(C^1)$ is equal to $A(1-p)$. We thus have to integrate over the range of $p$ values to find the area associated with the CS normalization prior,
\begin{equation}
N\int_0^{0.995} \frac{1}{1-p}~dp \approx 5.29832N.
\end{equation}

Finally, CS has a third parameter, $\alpha$. For each value of $1-p$, we allow this to range between zero (pure exponential) and $\alpha(p)$, where $\alpha(p)$ is set to give a $1/e$ cutoff at $200$ repeats. As an example, $\alpha(0.995)$ is zero; if $\alpha$ were greater than zero, the overall function would have an exponential cutoff longer than $200$ repeats. Given these, the area factor for nEXP is $0.995^n$, and for CS is it
\begin{equation}
\int^{0.995}_0 \alpha(p)~dp \approx 1.28841.
\end{equation}

Putting together all these area factors, we can then pre-compute $-\log{A}$, equal to $\log{P(\vec{w}_\textrm{max}|M)}$, a constant independent of $\vec{w}$. For the {\tt George\_W.\_Bush} article, for example, we have $-\log{A}$ equal to $-12.6$ for the CS case, and $-10.3$ (1EXP), $-18.7$ (2EXP), $-27.4$ (3EXP). Note that prior areas are not directly comparable between different models; ``change of units'' (\emph{e.g.}, working in terms of $P(RC^kR)$ vs. $N(RC^kR)$) will scale $A$. This scaling, however, is directly compensated for by the Hessian determinant term.

Together with the max log-likelihood, the determinant of the Hessian, and the $+k\log{2\pi}$, these are sufficient to compute the (Gaussian approximation to) the relative log-Evidence for the two model classes $\Delta\mathcal{E}$, reported in Table~\ref{evidence-long} (Table~2 in the Main Paper). In general, the highest evidence member of the nEXP class is either 3EXP or 4EXP. Table~\ref{evidence-long} gives the results for the top thirty most-edited pages.

\begin{table}
\begin{tabular}{l|l|l|l|l}
sig. & page name & history length & $\Delta\mathcal{E}$ & collective state index \\  
 & & & CS \emph{vs.} nEXP & $\alpha$ \\ \hline 
$<10^{-8}$ & {\tt George\_W.\_Bush} & 45,220 & 18.5 & $0.576\pm0.005$\\ 
$<10^{-6}$ & {\tt World\_War\_I} & 14,808 & 15.9 & $0.521\pm0.009$ \\ 
 & {\tt Islam} & 18,054 & 14.9 & $0.592\pm0.007$ \\ 
$<10^{-5}$ & {\tt Iraq\_War} & 15,143 & 12.8 & $0.60\pm0.01$ \\ 
 & {\tt Scientology} & 14,584 & 12.2 & $0.595\pm0.009$ \\ 
 & {\tt United\_States} & 31,919 & 12.3 & $0.545\pm0.006$ \\ 
 & {\tt Global\_warming} & 19,541 & 12.1 & $0.602\pm0.008$ \\ 
$<10^{-4}$ & {\tt Australia} & 13,815 & 11.4 & $0.679\pm0.009$ \\ 
 & {\tt Wikipedia} & 31,927 & 11.3 & $0.638\pm0.006$ \\ 
 & {\tt September\_11\_attacks} & 17,253 & 11.3 & $0.530\pm0.008$ \\ 
 & {\tt Gaza\_War} & 14,764 & 11.3 & $0.45\pm0.01$\\ 
 & {\tt Israel} & 16,319 & 11.1 & $0.523\pm0.008$ \\ 
 & {\tt Super\_Smash\_Bros.\_Br} & 15,343 & 11.1 & $0.451\pm0.008$ \\ 
 & {\tt Turkey} & 14,384 & 11.1 & $0.501\pm0.009$ \\ 
 & {\tt List\_of\_Omnitrix\_ali} & 16,263 & 10.6 &  $0.450\pm0.008$ \\ 
 & {\tt Michael\_Jackson} & 26,977 & 10.4 &  $0.572\pm0.007$ \\ 
 & {\tt Canada} & 17,670 & 9.4 & $0.632\pm0.008$ \\ 
 & {\tt Blink-182} & 14,419 & 9.4 & $0.461\pm0.009$ \\ 
$<10^{-3}$ & {\tt 2006\_Lebanon\_War} & 19,656 & 9.1 & $0.486\pm0.009$  \\ 
 & {\tt Blackout\_(Britney\_Sp} & 15,714 & 7.9 & $0.348\pm0.009$ \\ 
 & {\tt Deaths\_in\_2009} & 20,902 & 7.7 & $0.416\pm0.009$ \\ 
$<10^{-2}$ & {\tt Heroes\_(TV\_series)} & 14,060 & 6.6 & $0.353\pm0.009$ \\ 
 & {\tt Xbox\_360} & 16,598 & 6.4 & $0.498\pm0.009$ \\ 
 & {\tt Lost\_(TV\_series)} & 14,714 & 5.1 & $0.387\pm0.008$\\ 
 & {\tt Paul\_McCartney} & 16,649 & 4.7 & $0.72\pm0.01$ \\ 
(no det.) & {\tt Eminem} & 17,417 & 4.3 & --- \\ 
 & {\tt Pink\_Floyd} & 15,730 & 2.9 & --- \\ 
 & {\tt Deaths\_in\_2006} & 14,072 & 0.8 & --- \\ 
$>10^{4}$ & {\tt Deaths\_in\_2007} & 18,215 & -11.5 & --- \\ 
$>10^{7}$ & {\tt Deaths\_in\_2008} & 19,072 & -17.5 & --- \\ \hline 
\end{tabular}
\caption{log-Evidence ratios for the thirty most-edited pages on Wikipedia. Computed from data last accessed 15 July 2013.}
 \label{evidence-long}
\end{table}
\clearpage

\begin{thebibliography}{10}
\providecommand{\url}[1]{\texttt{#1}}
\providecommand{\urlprefix}{URL }
\expandafter\ifx\csname urlstyle\endcsname\relax
  \providecommand{\doi}[1]{doi:\discretionary{}{}{}#1}\else
  \providecommand{\doi}{doi:\discretionary{}{}{}\begingroup
  \urlstyle{rm}\Url}\fi
\providecommand{\bibAnnoteFile}[1]{%
  \IfFileExists{#1}{\begin{quotation}\noindent\textsc{Key:} #1\\
  \textsc{Annotation:}\ \input{#1}\end{quotation}}{}}
\providecommand{\bibAnnote}[2]{%
  \begin{quotation}\noindent\textsc{Key:} #1\\
  \textsc{Annotation:}\ #2\end{quotation}}
\providecommand{\eprint}[2][]{\url{#2}}

\bibitem{Hayek:1945wp}
Hayek FA (1945) {The use of knowledge in society}.
\newblock The American Economic Review XXXV: 519--530.
\input{Hayek:1945wp}

\bibitem{cottrell:1993cg}
Cottrell A, Cockshott WP (1993) {Calculation, complexity and planning: the
  Socialist Calculation debate once again}.
\newblock Quantitative Finance 5: 73--112.
\input{cottrell:1993cg}

\bibitem{lansing2007priests}
Lansing JS, Clark W (2007) Priests and Programmers: Technologies of Power in
  the Engineered Landscape of Bali.
\newblock Princeton University Press.
\input{lansing2007priests}

\bibitem{campbell1980american}
Campbell A, Converse PE, Miller WE, Stokes DE (1980) The American Voter.
\newblock Midway Reprint Series. University of Chicago Press.
\input{campbell1980american}

\bibitem{carpini1997americans}
Carpini M, Keeter S (1997) What Americans Know About Politics and Why It
  Matters.
\newblock Yale University Press.
\input{carpini1997americans}

\bibitem{DeCanio:1998fq}
DeCanio SJ, Watkins WE (1998) {Information processing and organizational
  structure}.
\newblock Journal of Economic Behavior and Organization 36: 275--294.
\input{DeCanio:1998fq}

\bibitem{Gluckman:1963vg}
Gluckman M (1963) {Gossip and Scandal}.
\newblock Current anthropology 4: 307--316.
\input{Gluckman:1963vg}

\bibitem{Gluckman:1968ul}
Gluckman M (1968) {Psychological, sociological and anthropological explanations
  of witchcraft and gossip: a clarification}.
\newblock Man 3: 20--34.
\input{Gluckman:1968ul}

\bibitem{chomsky1969aspects}
Chomsky N (1969) Aspects of the Theory of Syntax.
\newblock MIT Press.
\newblock Ch.~1.8.
\input{chomsky1969aspects}

\bibitem{Zimmerer:2010ix}
Zimmerer VC, Cowell PE, Varley RA (2010) {Individual behavior in learning of an
  artificial grammar}.
\newblock Memory {\&} Cognition 39: 491--501.
\input{Zimmerer:2010ix}

\bibitem{2011PLSCB...7E0011J}
Jin DZ, Kozhevnikov AA (2011) {A compact statistical model of the song syntax
  in Bengalese finch}.
\newblock PLoS computational biology 7: e1001108.
\input{2011PLSCB...7E0011J}

\bibitem{tenCate:2012ki}
ten Cate C, Okanoya K (2012) {Revisiting the syntactic abilities of non-human
  animals: natural vocalizations and artificial grammar learning}.
\newblock Philosophical transactions of the Royal Society of London Series B,
  Biological sciences 367: 1984--1994.
\input{tenCate:2012ki}

\bibitem{moore2011nature}
Moore C, Mertens S (2011) The Nature of Computation.
\newblock Oxford University Press.
\input{moore2011nature}

\bibitem{de2010grammatical}
de~la Higuera C (2010) Grammatical Inference: Learning Automata and Grammars.
\newblock Cambridge University Press.
\input{de2010grammatical}

\bibitem{wade-bayes}
Wade PR (2000) {Bayesian Methods in Conservation Biology}.
\newblock Conservation Biology 14: 1308--1316.
\input{wade-bayes}

\bibitem{suchard-bayes}
Suchard MA, Weiss RE, Dorman KS, Sinsheimer JS (2002) {Oh Brother, Where Art
  Thou? A Bayes Factor Test for Recombination with Uncertain Heritage}.
\newblock Systematic Biology 51: 715-728.
\input{suchard-bayes}

\bibitem{2011PhRvD..83d3505M}
Mortonson MJ, Peiris HV, Easther R (2011) {Bayesian analysis of inflation:
  Parameter estimation for single field models}.
\newblock Physical Review D 83: 43505.
\input{2011PhRvD..83d3505M}

\bibitem{2012PhRvD..85j3533E}
Easther R, Peiris HV (2012) {Bayesian analysis of inflation. II. Model
  selection and constraints on reheating}.
\newblock Physical Review D 85: 103533.
\input{2012PhRvD..85j3533E}

\bibitem{2012PhRvD..86b3505N}
Nore{\~n}a J, Wagner C, Verde L, Peiris HV, Easther R (2012) {Bayesian analysis
  of inflation. III. Slow roll reconstruction using model selection}.
\newblock Physical Review D 86: 23505.
\input{2012PhRvD..86b3505N}

\bibitem{powell-bayes}
Powell BA, Tzirakis K, Kinney WH (2009) Tensors, non-gaussianities, and the
  future of potential reconstruction.
\newblock Journal of Cosmology and Astroparticle Physics 2009: 019.
\input{powell-bayes}

\bibitem{2012arXiv1209.2024P}
{Powell} BA (2012) {Scalar runnings and a test of slow roll from CMB
  distortions}.
\newblock  arXiv:12092024 .
\input{2012arXiv1209.2024P}

\bibitem{kass-bayes}
Kass RE, Raftery AE (1995) {Bayes Factors}.
\newblock Journal of the American Statistical Association 90: 773-795.
\input{kass-bayes}

\bibitem{mackay2003information}
MacKay DJC (2003) Information Theory, Inference and Learning Algorithms.
\newblock Cambridge University Press.
\newblock Ch.~28.
\input{mackay2003information}

\bibitem{Suh:2007cb}
Suh B, Chi EH, Pendleton BA, Kittur A (2007) {Us vs. Them: Understanding Social
  Dynamics in Wikipedia with Revert Graph Visualizations}.
\newblock In: IEEE Symposium on Visual Analytics Science and Technology. IEEE,
  pp. 163--170.
\input{Suh:2007cb}

\bibitem{Kittur:2007ey}
Kittur A, Suh B, Pendleton BA, Chi EH (2007) {He says, she says: conflict and
  coordination in Wikipedia}.
\newblock In: Proceedings of the SIGCHI conference on Human factors in
  computing systems. ACM Press, pp. 453--462.
\input{Kittur:2007ey}

\bibitem{Brandes:2008fo}
Brandes U, Lerner J (2008) {Visual analysis of controversy in user-generated
  encyclopedias⋆}.
\newblock Information Visualization 7: 34--48.
\input{Brandes:2008fo}

\bibitem{Kittur:2010ic}
Kittur A, Kraut RE (2010) {Beyond Wikipedia: Coordination and Conflict in
  Online Production Groups}.
\newblock In: Proceedings of the 2010 ACM conference on Computer supported
  Cooperative Work. Savannah, GA,: ACM Press, p. 215.
\input{Kittur:2010ic}

\bibitem{DeDreu:2003gz}
De~Dreu CKW, Weingart LR (2003) {Task versus relationship conflict, team
  performance, and team member satisfaction: A. meta-analysis}.
\newblock Journal of Applied Psychology 88: 741--749.
\input{DeDreu:2003gz}

\bibitem{reagle}
Reagle J (2010) {Good Faith Collaboration: The Culture of Wikipedia}.
\newblock History and Foundations of Information Science Series. MIT Press.
\input{reagle}

\bibitem{cosley2006helping}
Cosley D (2006) Helping Hands: Design for Member-maintained Online Communities.
\newblock University of Minnesota.
\input{cosley2006helping}

\bibitem{Boyd:1989fz}
Boyd R (1989) {Mistakes allow evolutionary stability in the repeated prisoner's
  dilemma game}.
\newblock Journal of Theoretical Biology 136: 47--56.
\input{Boyd:1989fz}

\bibitem{Raub:1990us}
Raub W, Weesie J (1990) {Reputation and Efficiency in Social Interactions: An
  Example of Network Effects}.
\newblock American Journal of Sociology 96: 626--654.
\input{Raub:1990us}

\bibitem{Bendor:2001ku}
Bendor J, Swistak P (2001) {The Evolution of Norms}.
\newblock American Journal of Sociology 106: 1493--1545.
\input{Bendor:2001ku}

\bibitem{Ostrom:2000vi}
Ostrom E (2000) {Collective action and the evolution of social norms}.
\newblock The Journal of Economic Perspectives 14: 137--158.
\input{Ostrom:2000vi}

\bibitem{bowles2011cooperative}
Bowles S, Gintis H (2011) A Cooperative Species: Human Reciprocity and Its
  Evolution.
\newblock Princeton University Press.
\input{bowles2011cooperative}

\bibitem{linke2012space}
Linke AM, Witmer FD, O'Loughlin J (2012) {Space-time granger analysis of the
  war in Iraq: A study of coalition and insurgent action-reaction}.
\newblock International Interactions 38: 402--425.
\input{linke2012space}

\bibitem{jackendoff2007language}
Jackendoff RS (2007) Language, Consciousness, Culture: Essays on Mental
  Structure.
\newblock Jean Nicod Lectures. {MIT} Press.
\input{jackendoff2007language}

\bibitem{Hauser:2002p20637}
Hauser MD, Chomsky N, Fitch T (2002) {The Faculty of Language: What Is It, Who
  Has It, and How Did It Evolve?}
\newblock Science 298: 1569--1579.
\input{Hauser:2002p20637}

\bibitem{Stoop:2012ux}
Stoop R, N{\"u}esch P, Stoop RL, Bunimovich L (2012) {Fly out-smarts man}.
\newblock  arXiv:12025913 .
\input{Stoop:2012ux}

\bibitem{Crutchfield199411}
Crutchfield JP (1994) The calculi of emergence: computation, dynamics and
  induction.
\newblock Physica D: Nonlinear Phenomena 75: 11--54.
\input{Crutchfield199411}

\bibitem{DeDeo:2011p19975}
DeDeo S (2011) {Effective theories for circuits and automata}.
\newblock Chaos 21: 7106.
\input{DeDeo:2011p19975}

\bibitem{DeDeo:2010p18133}
DeDeo S, Krakauer DC, Flack JC (2010) {Inductive Game Theory and the Dynamics
  of Animal Conflict}.
\newblock PLoS computational biology 6: e1000782.
\input{DeDeo:2010p18133}

\bibitem{DeDeo:2011p18663}
DeDeo S, Krakauer DC, Flack JC (2011) Evidence of strategic periodicities in
  collective conflict dynamics.
\newblock Journal of The Royal Society Interface 8: 1260--1273.
\input{DeDeo:2011p18663}

\bibitem{Flack:2011p20179}
Flack JC, Krakauer DC (2011) {Challenges for complexity measures: A perspective
  from social dynamics and collective social computation}.
\newblock Chaos 21: 7108.
\input{Flack:2011p20179}

\bibitem{hartmanis1966algebraic}
Hartmanis J, Stearns RE (1966) Algebraic structure theory of sequential
  machines.
\newblock Prentice-Hall international series in applied mathematics.
  Prentice-Hall.
\input{hartmanis1966algebraic}

\bibitem{berman1987nonnegative}
Berman A, Plemmons RJ (1987) Nonnegative Matrices in the Mathematical Sciences.
\newblock Classics in Applied Mathematics. Society for Industrial and Applied
  Mathematics.
\newblock Ch.~2.3.
\input{berman1987nonnegative}

\bibitem{seneta2006non}
Seneta E (2006) Non-negative Matrices and Markov Chains.
\newblock Springer Series in Statistics. Springer.
\newblock Ch.~1.2.
\input{seneta2006non}

\bibitem{meyer2000matrix}
Meyer C (2000) {Matrix Analysis and Applied Linear Algebra}.
\newblock SIAM Press.
\newblock Ch.~8.
\input{meyer2000matrix}

\bibitem{andrieux}
{Andrieux} D (2011) {Spectral Signature of Nonequilibrium Conditions}.
\newblock  arXiv:11032243 .
\input{andrieux}

\bibitem{pippenger1997theories}
Pippenger N (1997) Theories of Computability.
\newblock Cambridge University Press.
\input{pippenger1997theories}

\bibitem{Berwick:2011jh}
Berwick RC, Okanoya K, Beckers GJL, Bolhuis JJ (2011) {Songs to syntax: the
  linguistics of birdsong}.
\newblock Trends in Cognitive Sciences 15: 113--121.
\input{Berwick:2011jh}

\bibitem{languagelog}
Liberman M (2011).
\newblock Finch linguistics.
\newblock In \emph{Language Log}, \\
  \href{http://languagelog.ldc.upenn.edu/nll/?p= 3261}{{\tt
  http://languagelog.ldc.upenn.edu/nll/?p=3261}}.
\newblock Informal analysis of data supplied by Ofer Tchernickovski and Dina
  Lipkind. Last accessed 15 August 2013.
\input{languagelog}

\bibitem{Yasseri:2012ga}
Yasseri T, Sumi R, Rung A, Kornai A, Kert{\'e}sz J (2012) {Dynamics of
  conflicts in Wikipedia}.
\newblock PLoS ONE 7: e38869.
\input{Yasseri:2012ga}

\bibitem{Kass:1995eh}
Kass RE, Raftery AE (1995) {Bayes Factors}.
\newblock Journal of the American Statistical Association 90: 773--795.
\input{Kass:1995eh}

\bibitem{Liddle:2004cz}
Liddle AR (2004) {How many cosmological parameters?}
\newblock Monthly Notices of the Royal Astronomical Society 351: L49--L53.
\input{Liddle:2004cz}

\end{thebibliography}

\end{document}